\documentclass[aps,prl,twocolumn,groupedaddress]{revtex4}
\providecommand{\E}{e}
\providecommand{\I}{i}
\providecommand{\D}{\mathrm{d}}
\renewcommand{\vec}{\mathbf}

\usepackage{makeidx}       % allows index generation
\usepackage{graphicx}        % standard LaTeX graphics tool
\begin{document}

\title{Current-Induced Effects in Nanoscale Conductors}
\author{Neil Bushong}
\email{bushong@physics.ucsd.edu}
\author{Massimiliano Di Ventra}
\email{diventra@physics.ucsd.edu}

\affiliation{Department of Physics, University of California, San Diego,
  La Jolla, CA 92093-0319}

\begin{abstract}
  \index{current-induced effects} We present an overview of
  current-induced effects in nanoscale conductors with emphasis on
  their description at the atomic level. In particular, we discuss
  steady-state current fluctuations, current-induced forces, inelastic
  scattering and local heating. All of these properties are calculated
  in terms of single-particle wavefunctions computed using a
  scattering approach within the static density-functional theory of
  many-electron systems. Examples of current-induced effects in atomic
  and molecular wires will be given and comparison with experimental
  results will be provided when available.
\end{abstract}

\maketitle

\section{Current Through a Nanoscale Junction}\label{Ctanj}
\index{current!through a nanoscale junction}
\index{nanoscale junction}

Transport of electrical charge across a nanoscale junction is
accompanied by many effects, such as fluctuations of the average
current; transfer of energy between electrons and ions and consequent
heating of the junction; and forces on ions due to current-induced
variations of the electronic distribution~\cite{diventrabook:04}.  In
this work we will discuss these effects separately, and we will
focus on their description at the atomic level. It is, however,
important to realize that there has to be a (yet unknown) relation
between these different properties. Such relation would constitute an
important contribution to our understanding of transport in nanoscale
systems.

The static scattering approach to electrical conduction will be the
underlying theme of this review. However, we point out that novel
time-dependent formulations of the transport problem may lead us to a
better understanding of these effects~\cite{diventra:04:jpcm,
  horsfield:04jpcm16-3609}.  In what follows we picture a nanoscale
junction as formed by two semi-infinite electrodes held a fixed
distance apart, with a nanoscale object bridging the gap between them.
The nanoscale object could be a single atom, a chain of atoms, a
molecule, or any system with nanoscale dimensions~\cite{ohnishi:98,
  yazdani:96, rodrigues:01, metzger:97, zhou:97, reed:97, datta:97,
  gimzewski:99, chen:99, reed:01, lang:00, diventra:02prb65,
  tomfohr:05}.  We then consider the problem of DC current flow from
one electrode to the other as the result of an applied bias
$V_\mathrm{B}$.  If we take the left electrode to be positively
biased, electrons will flow from the right electrode to the left one.
A buildup of negative charge will be present on the surface of the
right electrode within a screening length of the electrode surface.
Similarly, a buildup of positive charge will exist on the surface of
the left electrode due to a corresponding depletion of
electrons~\cite{lang:00, diventra:02prb65}.  We will assume that, far
away from the junction, the electrons in each electrode are in local
thermal equilibrium and their statistics are described by the
Fermi-Dirac distribution, so that $V_\mathrm{B} = E_\mathrm{FR} -
E_\mathrm{FL}$, where $E_\mathrm{FL(R)}$ is the chemical potential
deep in the left (right) electrode.  (Here, as in the rest of this
work, we use atomic units.)

We calculate the transport properties of this system by expanding the
stationary states of the Hamiltonian into a set of left- and
right-moving waves.  We then sum each left- and right-moving state,
weighting them with a Fermi function according to their energy.  The
stationary scattering states of the bare electrodes have the form
\begin{equation}
\Psi_{E\vec{K}_\parallel}^0(\vec{r}) =
\E^{ \I\vec{K}_\parallel \cdot \vec{Y} } u_{E\vec{K}_\parallel}(z),
\end{equation}
with the following boundary conditions~\cite{messiah:99}:
\begin{equation}
u_{E\vec{K}_\parallel}(z)
= (2\pi)^{3/2} k_\mathrm{R}^{-1/2}
\times \left\{ %
\begin{array}{ll}
 \E^{- \I k_\mathrm{R}z}
 + R\E^{\I k_\mathrm{R}z}, & z \rightarrow \infty \\
 T \E^{- \I k_\mathrm{L}z}, & z \rightarrow -\infty. %
\end{array} \right.
\end{equation}
$\vec{K}_\parallel$ is the electron momentum in the plane parallel to
the electrode surfaces.  We have defined $\frac{1}{2} k_\mathrm{R}^2 =
E - \frac{1}{2}|\vec{K}_\parallel|^2 - v_{eff}(\infty)$, and
$\frac{1}{2} k_\mathrm{L}^2 = E - \frac{1}{2}| \vec{K}_\parallel|^2 -
v_{eff}(-\infty)$.  $\vec{Y}$ is the component of the position vector
in a plane parallel to the electrode surfaces, and $z$ is the
coordinate perpendicular to them. $v_{eff}(\pm \infty)$ is the bottom
of the electronic energy band deep within the right/left electrode.
The wavefunctions $\Psi^0$ satisfy the continuum normalization
condition
\begin{equation}\label{NormCond}
\int \D ^3\vec{r}[\Psi_{E'\vec{K}_\parallel'}^0(\vec{r})]^*
\Psi_{E\vec{K}_\parallel}^0(\vec{r}) =
\delta(E'-E)\delta(\vec{K_\parallel}' - \vec{K}_\parallel).
\end{equation}

Since the details of the electrodes are not important up to the
interface with the sample, we represent them using a
uniform-background (jellium) model~\cite{lang:00, diventra:02prb65}.
The potential $V$ the electrons experience when they scatter through
the nanojunction is~\footnote{For the detailed implementation of the
  approach outlined in this work see the original
  papers~\cite{lang:00, diventra:02prb65}.}
\begin{eqnarray}\label{PotDens}
V(\vec{r},\vec{r}') & = & v_\mathrm{ps}(\vec{r},\vec{r}') + 
\bigg[ v_\mathrm{xc}[n(\vec{r})]
- v_\mathrm{xc}[n^0(\vec{r})] \nonumber \\
& & + \int \D ^3\vec{r}''
\frac{\delta n(\vec{r}'')}{|\vec{r} - \vec{r}''|} \bigg]
\delta(\vec{r} - \vec{r}').
\end{eqnarray}
The term $v_\mathrm{ps}$ is the electron-ion interaction potential
that we represent with (nonlocal) pseudopotentials; $v_\mathrm{xc}$ is
the exchange-correlation potential computed using the local-density
approximation to density functional theory (DFT)~\footnote{One
  possible choice of exchange-correlation functional is the one given
  in \cite{ceperley:80}, as parametrized in \cite{perdew:81}.  For an
  extended discussion of the local density approximation, see
  reference~\cite{dicarlo:05}.}; $n^0(\vec{r})$ is the electronic
density for the pair of biased bare electrodes; $n(\vec{r})$ is the
electronic density for the total system, and $\delta n(\vec{r})$ is
their difference. We are implicitely assuming that static DFT gives a
reasonable account of the scattering properties of a nanoscale system,
at least in linear response. While this may be true for metallic
junctions it is not obvious for molecular junctions~\cite{sai:04}.
However, in linear response and far from the resonant regime, we
expect basic physical trends for these systems to be reproduced well
by static DFT~\cite{sai:04}.

The full Hamiltonian of the system is $H = H_0 + V$, where $H_0$ is
the Hamiltonian due to the bare biased electrodes, and $V$ is the
scattering potential.  Our next task is to find the self-consistent
solutions to the equation $H\Psi_E = E\Psi_E$, which we can put into
the Lippmann-Schwinger form:
\begin{eqnarray}\label{Lipmann-Schwinger}
\lefteqn{
\Psi_{E\vec{K}_\parallel}(\vec{r}) =
\Psi_{E\vec{K}_\parallel}^0 (\vec{r}) } \nonumber \\
& & + \int \D ^3\vec{r}' \D ^3\vec{r}'' 
G_E^0(\vec{r},\vec{r}') V(\vec{r}',\vec{r}'') 
\Psi_{E\vec{K}_\parallel} (\vec{r}'').
\end{eqnarray}
The quantity $G_E^0$ is the Green's function for the bare electrodes,
and needs to be calculated for each energy $E$.

Lastly, the total density of the system is
\begin{equation}\label{density}
n(\vec{r}) = 2\sum_i |\Psi_i(\vec{r})|^2 + 
2 
\int\! \D E\int\! \D ^2\vec{K}_\parallel 
|\Psi_{E\vec{K}_\parallel} (\vec{r})|^2,
\end{equation}
where we have included a factor of 2 due to spin degeneracy.  The
$\Psi_i$'s are the bound states of $H$, if any exist.  They can be
calculated by direct diagonalization of the full Hamiltonian $H$.  In
order to find a self-consistent solution for the density, equations
(\ref{PotDens}), (\ref{Lipmann-Schwinger}) and (\ref{density}) are
solved iteratively.

One has the freedom to choose a basis set to represent the wave
functions. Due to the non-variational nature of the electrical
current, the issue of which basis set to use in a given calculation is
far from trivial~\cite{yang:02}.  In the examples that follow, plane
waves have been used as basis set~\footnote{For other examples of
  calculations using different basis sets, see Section 2.2 of
  reference~\cite{stokbro:05}, and the references therein.}, which
allows easy testing of the convergence of the
results~\cite{yang:02}\index{plane wave basis set}.

Once the wavefunctions have been calculated self-consistently, the
total electric current density (at zero temperature) is given by
\begin{equation}
\vec{j}(\vec{r}) = -2
\int_{E_\mathrm{FL}}^{E_\mathrm{FR}} \! \D E 
\int\! \D ^2\vec{K}_\parallel 
\mathrm{Im}\{[\Psi_{E\vec{K}_\parallel}(\vec{r})]^* 
\nabla\Psi_{E\vec{K}_\parallel}(\vec{r})\}.
\end{equation}
The quantity we are interested in is the extra current $\delta J$ due
to presence of the nanojunction~\cite{lang:92, lang:95prb51errat,
  lang:94, lang:95prb52} (two semi-infinite contacts have infinite
surface area, and thus pass infinite current)
\begin{equation}\label{current}
\delta J = \int \D ^2\vec{Y} \hat{z} \cdot
[ \vec{j}(\vec{r}) - \vec{j}^0(\vec{r}) ],
\end{equation}
where $\vec{j}$ is the electric current density, and we have defined
$\vec{j}^0$ to be the current density in the absence of the
nanojunction. The current defined in equation (\ref{current}) is the
average current that flows across the nanojunction. Fluctuations with
respect to that average are expected and will be described later in
this work.

It is worth noting that, in contrast to references~\cite{jortner:05,
  hanggi:05, stokbro:05, dicarlo:05}, we do not calculate the
electrical current using any of the Landauer
formulas~\cite{landauer:92, landauer:87}\index{Landauer-B{\"u}ttiker
  formulation}; rather, the current is calculated as the expectation
value of the current operator over single-particle states. Clearly, it
can be proven that when the transmission probabilities are extracted
from the single-particle scattering wavefunctions, we recover from
equation (\ref{current}) the conventional two-probe Landauer
formula~\cite{lagerqvist:04}.

\section{Current-Induced Forces}\label{C-if}%
\index{current-induced forces}
\index{force!current-induced forces}

Now that we have computed the stationary scattering states and
corresponding current we can look at one of the effects induced by
electron flow. We first focus on current-induced forces, i.e. the
phenomenon by which atoms in a current-carrying wire are subject to
forces due to the local change in self-consistent electronic
distribution~\cite{diventra:00prb61, seideman:03}.  There are several
open questions related to current-induced forces, the most notable of
which is their conservative nature~\cite{diventra:04}.  Here we will
focus on their dependence on some microscopic properties. We will show
that these forces are a non-linear function of the junction
properties, such as its current density, charge density and length.

\begin{figure}[!bhp]
  \centering
  \includegraphics[width=.4\textwidth,clip]{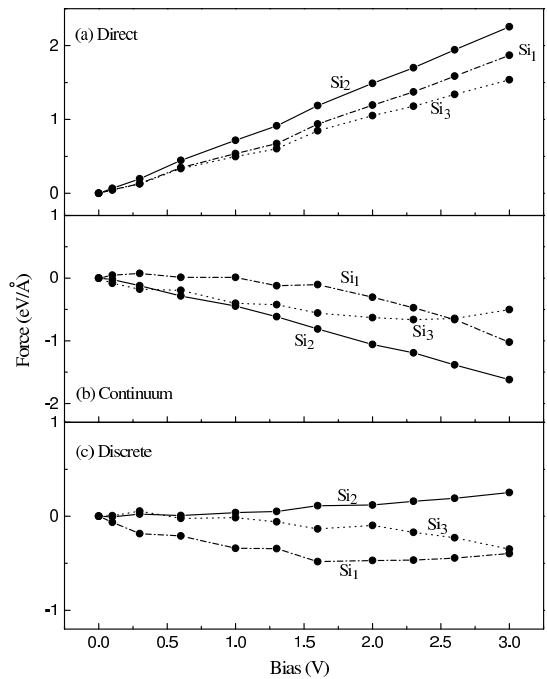}
  \caption{Different contributions to the total force for each
    of the three atoms composing a silicon wire, as a function of
    bias. See text for details.  Reprinted figure with permission from
    \protect\cite{yang:03}.  Copyright 2003 by the American Physical
    Society.}
  \label{forceplot}
\end{figure}

Let us first define forces in a current-carrying wire. We first note
that in this non-equilibrium problem the usual Hellman-Feynmann
theorem \index{Hellman-Feynan theorem} is not
valid~\cite{diventra:00prb61,diventra:04}.  A meaningful definition of
force on ions can then be obtained by either the classical limit of
Ehrenfest's theorem \index{Ehrenfest's theorem} applied to the rate of
change of ionic momentum~\cite{diventra:00prb61} or from the
Euler-Lagrange equation of motion for the classical
ions~\cite{todorov:01}.  Both approaches yield the same expression for
the force on an ion, with position ${\bf R}$, due to the
self-consistent electronic density $\rho({\bf r})$ under current
flow~\cite{todorov:01},
\begin{equation}
{\bf F}= - 
\int \D{\bf r}
\frac{\partial v_\mathrm{ps}}{\partial {\bf R}} \rho({\bf r}),
\label{equation}
\end{equation}
which can be equivalently written as~\cite{diventra:00prb61}
\begin{equation}
{\bf F}= - 
\sum_i \bigg\langle \psi_i\left|
\frac{\partial v_\mathrm{ps}}{\partial{\bf R}}\right|\psi_i\bigg\rangle
- 
\lim_{{\Delta}\rightarrow 0} \int_{\sigma}dE
\bigg\langle \psi_{\Delta}\left|
\frac{\partial v_\mathrm{ps}}{\partial {\bf R}}
\right|\psi_{\Delta}\bigg\rangle.
\label{equation1}
\end{equation}
The first term on the RHS of Eq.~(\ref{equation1}) is similar to the
usual Hellmann-Feynman contribution to the force due to localized
electronic states $|\psi_i\rangle$. The second term is the
contribution due to continuum states~\cite{diventra:00prb61}. The
wavefunctions $|\psi_{\Delta}\rangle$ are eigendifferentials for each
energy interval $\Delta$ in the continuum
$\sigma$~\cite{diventra:00prb61}.  Finally, an additional contribution
from ion-ion interactions needs to be taken into account.

For convenience we can separate the total force into two
contributions. We define ``direct force'' to be the portion of the
total force due to the charge distribution of the biased bare
electrodes. We then call the remaining term the ``electron wind
force''\footnote{Note that this definition of ``direct force'', and
  corresponding ``wind force'', is not universal in literature.  For a
  discussion of the various definitions, see reference
  \cite{sorbello:98}.}\index{wind force}\index{direct force}.  There
exists an inelastic portion of the electron wind force due to energy
transfer from electrons to ions, in the form of phonon excitations;
however, this contribution is generally negligible compared to the
elastic one and will be neglected here~\cite{yang:03}.

We note that atomic relaxations induced by current flow do not have a
large effect on the absolute value of the current that passes through
the junction, even in the limit of high voltages and current
densities~\cite{diventra:00prb61, yang:03, diventra:00prl88,
  diventra:02cp281}.  While this result says nothing about the
mechanical stability of current-carrying wires under bias, it allows
us to study several transport properties assuming atomic positions
fixed at their zero bias value.

\begin{figure}[!bhp]
  \centering
  \includegraphics[width=.4\textwidth,clip]{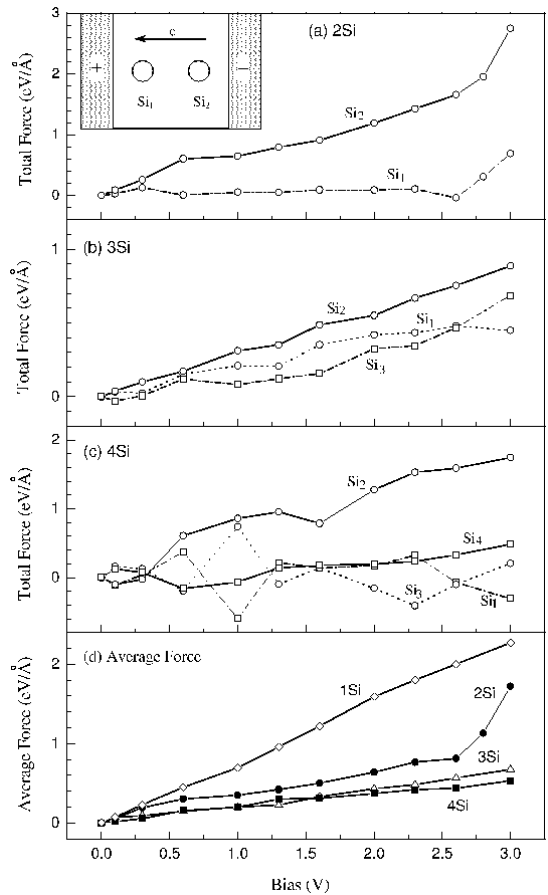}
  \caption{Total force on each atom in atomic wires consisting of
    (a)two, (b)three, and (c) four silicon atoms.  The inset shows a
    schematic of the wire that is composed of two silicon atoms.  (d)
    The average force on each wire.  Reprinted figure with permission
    from \protect\cite{yang:03}.  Copyright 2003 by the American
    Physical Society.}
  \label{diffnumbers}
\end{figure}

Let us now look at the dependence of these forces on microscopic
details.  As an example, in Figure \ref{forceplot} we consider a
nanoscopic wire composed of three silicon atoms between two bulk
electrodes.  We keep atomic positions fixed at their zero bias values.
For each atom in the junction, we plot the direct force, the electron
wind force due to the continuum states, and the electron wind force
due to the discrete part of the spectrum.

The direct force is almost linear with applied bias, with the force on
the central atom being the largest.  Closer to the electrode surfaces,
the electrostatic potential ceases to be a linear function of
position, causing the force on the other two atoms to be slightly
smaller.  Note that the force on the central atom due to states in the
continuum is positive (i.e. pushes the atom to the left, in the same
direction as the electron flow).  However, the force on the central
atom due to states in the discrete spectrum is almost zero, even at
larger biases.  The reason for this behavior is that the amount of
charge localized around the central atom is almost constant in the
applied bias, and so the corresponding force is also
constant~\cite{yang:03}.

Figure \ref{diffnumbers}(a) through (c) shows the force on specific
atoms in silicon wires composed of varying numbers of atoms, while
Figure \ref{diffnumbers}(d) shows the average force on the total
wires.  It is difficult to extract overall trends from Figures
\ref{diffnumbers}(a) through (c), although it is interesting to note
that the second atom from the left (labeled Si$_2$) is consistently
the atom that experiences the greatest force.  Figure
\ref{diffnumbers}(d), however, follows an obvious trend: the average
force saturates with increasing number of atoms.  This comes from the
fact that, as the length of the wire increases, the boundary effects
due to the electrodes become less important. This result suggests that
longer wires are more difficult to break under current flow, an
important factor to consider in nanoscale devices.

\section{Shot Noise}\label{Sn}
\index{shot noise}
\index{noise!shot noise}

In this section we will discuss steady-state current fluctuations that
occur due to the quantization of charge. Shot noise is quite distinct
from (equilibrium) thermal noise and is generally unavoidable, even at
zero temperature~\cite{blanter:00}.

We will first derive an expression of shot noise in terms of
single-particle wavefunctions. In order to do so, let us write the
field operator for electrons $\hat{\Psi}$ as combination of a field
operator due to electrons incident from the left, and one due to
electrons incident from the right~\cite{chen:03prb67,lagerqvist:04}
\begin{equation}
\hat{\Psi} = \hat{\Psi}^\mathrm{L}
+ \hat{\Psi}^\mathrm{R} \label{fieldop}.
\end{equation}
We can further expand each of these field operators into the single
particle wave functions~(\ref{Lipmann-Schwinger}):
\begin{equation}
\hat{\Psi}^\mathrm{L(R)} = \sum_E  \E^{-\I \omega t}
a_E^\mathrm{L(R)}
\Psi_E^\mathrm{L(R)}(\vec{r},\vec{K}_\parallel).\label{atExpans}
\end{equation}
Again, we have used $\vec{K}_\parallel$ to denote the component of the
electron's incident momentum that is parallel to the surfaces of the
electrodes~\footnote{The presence of the nanojunction causes
  $\vec{K}_\parallel$ to cease to be a good quantum number; we still
  use it here as a ``label'' to enumerate scattering states.}.

The coefficients $a_E^\mathrm{L(R)}$ are the annihilation operators
for electrons incident from the left (right) reservoir.  These
operators satisfy the usual anticommutation relation
$\{a_E^\phi,a_{E'}^{\chi\dagger}\} = \delta_{\phi\chi}\delta(E-E')$,
where $\phi,\chi = \mathrm{R,L}$. We will again assume that the
electrons coming from the left (right) reservoir are in local thermal
equilibrium at a temperature $T_e$ far away from the junction, so that
their statistics are given by the Fermi-Dirac distribution function
$f_E^\mathrm{L(R)}$, i.e.
\begin{eqnarray}
\langle a_E^{\phi\dagger}a_{E'}^\chi \rangle & = &
\delta_{\phi\chi}\delta(E-E')
f_E^\phi \\
& = & \frac{\delta_{\phi\chi}\delta(E-E')}{
\E^{[E - E_{F\phi}]/k_\mathrm{B}T_\mathrm{e}}+1 }.
\end{eqnarray}

Using the field operator (\ref{fieldop}), we can define the current
operator
\begin{equation}
\hat{I}(z,t) =
-\I \int \D \vec{Y}\int \D \vec{K_\parallel}
(\hat{\Psi}^\dagger \partial_z\hat{\Psi} -
\partial_z\hat{\Psi}^\dagger\hat{\Psi}).
\end{equation}
In the limit of zero temperature, the Fermi distribution reduces to a
step function; if we again assume the chemical potential in the right
reservoir $E_\mathrm{FR}$ is higher than the chemical potential in the
left reservoir $E_\mathrm{FL}$, then at zero temperature the average
value of the current is just
\begin{equation} \label{currentop}
\langle \hat{I} \rangle = -\I \int_{E_\mathrm{FL}}^{E_\mathrm{FR}} \!
\D E \int\! \D \vec{Y} \int\!
\D \vec{K}_\parallel \tilde{I}_{E,E}^{R,R},
\end{equation}
where
\begin{equation} \label{itilde}
\tilde{I}_{E,E'}^{\phi,\chi} = (\Psi_E^\phi)^*\nabla\Psi_{E'}^\chi -
\nabla(\Psi_E^\phi)^*\Psi_{E'}^\chi,
\end{equation}
and $^*$ denotes complex conjugation.

We define shot noise as the Fourier transform of the electric current
autocorrelation function in the limit of zero frequency and zero
temperature~\cite{blanter:00,camalet:03}:
\begin{equation}
2\pi S(\omega) = \int\! \D t \, \E^{\I \omega t} \langle
\Delta\hat{I}(t)\Delta\hat{I}(0) \rangle,
\end{equation}
where the excess current operator $\Delta\hat{I}(t)$ is equal to
$\hat{I}(t) -\langle\hat{I} \rangle$.  The evaluation of this
expression is quite cumbersome and therefore we do not include it
here. We only mention that we need to evaluate terms of the form $
\langle \hat{A}_4 \hat{A}_3 \hat{A}_2 \hat{A}_1 \rangle$, where the
$\hat{A}_i$'s are raising and lowering operators. We can evaluate
these terms with the Bloch-De Dominicis theorem~\cite{kubo:92}:
\begin{eqnarray}
\langle \hat{A}_4 \hat{A}_3 \hat{A}_2 \hat{A}_1 \rangle
& = & \langle \hat{A}_4 \hat{A}_3 \rangle \langle \hat{A}_2 \hat{A}_1 \rangle
+ \eta \langle \hat{A}_4 \hat{A}_2 \rangle \langle \hat{A}_3 \hat{A}_1
\rangle \nonumber \\
& & + \eta^2 \langle \hat{A}_4 \hat{A}_1 \rangle \langle \hat{A}_3
\hat{A}_2 \rangle,
\end{eqnarray}
where $\eta = -1$ for fermions, and $+1$ for bosons.  The end result
is
\begin{eqnarray}\label{BigNoise}
S(\omega) & = & \! \sum_{\phi,\chi = \mathrm{L},\mathrm{R}} 
\int\!\! \D E f_{E+\omega}^\phi (1 - f_E^\chi)
\int\!\! \D \vec{Y}_1 \!
\int\!\! \D \vec{K}_1  \nonumber \\
& & \times \tilde{I}_{E+\omega,E}^{\phi\chi} \!
\int\!\! \D \vec{Y}_2 \!
\int\!\! \D \vec{K}_2 \tilde{I}_{E,E+\omega}^{\chi\phi}.
\end{eqnarray}

In the limit of zero frequency and zero temperature, equation
(\ref{BigNoise}) reduces to~\cite{chen:03prb67,lagerqvist:04}
\begin{equation}
S = \int_{E_\mathrm{FL}}^{E_\mathrm{FR}} \D E
\bigg|\int \D \vec{R}
\int \D \vec{K} \tilde{I}_{E,E}^{\mathrm{LR}} \bigg|^2.
\end{equation}
This is the desired expression relating shot noise to single-particle
wavefunctions~\footnote{As for the current, this expression can be
  reduced to the well-known result $S=\frac{V_B}{\pi} \sum_n
  T_n(1-T_n)$~\cite{buttiker:90} if the transmission probabilities
  $T_n$ of each mode are extracted from the scattering wavefunctions;
  see reference~\cite{lagerqvist:04}.}.

In the case of uncorrelated electrons, the magnitude of the shot noise
becomes $S_\mathrm{P}=2eI$,\index{Poisson limit}\index{shot
  noise!Poisson limit}\index{Schottky's value for the Poisson limit}
where $e$ is the electron charge, and $I$ is the dc current.  This
corresponds to the noise of a series of incident particles, the time
between the arrivals of which follows a Poissonian distribution
function.  For this reason, $S_\mathrm{P}$ is sometimes called the
Poisson value for the shot noise~\cite{blanter:00, schottky:18}.  In
general, however, the magnitude of the shot noise will be less than
the Poisson value.  A relative measure of noise is therefore the Fano
factor $F$, defined as the ratio between shot noise $S$ and the
Poisson limit $S_\mathrm{P}$.\index{Fano factor}

We are now ready to discuss an example of noise properties of a
nanoscale junction. We again look at the properties of a junction
formed by a short wire of silicon atoms between two bulk electrodes.
Other examples of noise in atomic-scale systems can be found in
references \cite{chen:03prb67} and \cite{lagerqvist:04}.

\begin{figure}
  \centering
  \includegraphics[width=.4\textwidth]{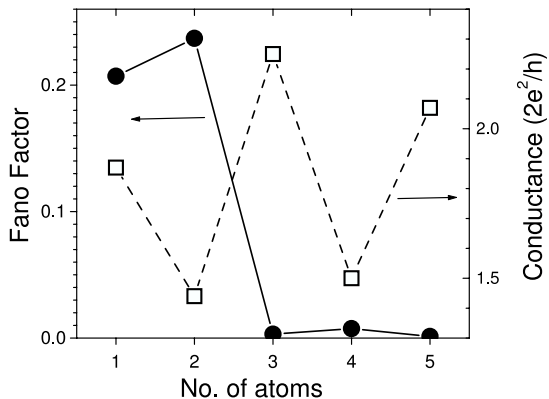}
  \caption{Fano factor and conductance for a nanojunction composed of
    different numbers of silicon atoms.  Reprinted figure with
    permission from \protect\cite{chen:03prb67}.  Copyright 2003 by
    the American Physical Society.}
  \label{FanoCond}
\end{figure}
Figure \ref{FanoCond} shows the results of conductance and Fano factor
for such a system for a bias of 0.01 V.  We first notice that the Fano
factor is strongly nonlinear as a function of bias. It is also
considerably enhanced for very short wires due to the large
contribution from the metal electrodes. In addition, the Fano factor
oscillates as a function of number of silicon atoms\index{Fano factor!oscillation with wire length}. The conductance shows a
similar but opposite oscillatory trend. Both the noise and conductance
oscillations are due to the fact that a Si wire made up of an even
number of atoms has fully-occupied $\pi$ orbitals, while a wire made
up of an odd number of atoms has a half-filled $\pi$ state at the
Fermi level~\cite{lang:97}.

Other interesting properties of shot noise in nanoscale systems
include its dependence on contact geometry and interwire
interactions~\cite{lagerqvist:04}.  We refer the reader to the
original papers for a discussion of these effects.

\section{Local Heating}
\index{local heating}
\index{heating}
\index{phonons}

Local heating occurs when electrons in a current-carrying wire
exchange energy with
phonons~\cite{chen:03:nl3,montgomery:02,montgomery:03, segal:00,
  troisi:03}.  Accordingly, there are four main processes that
contribute to local heating in the junction:
\begin{enumerate}
  
\item Cooling processes in which an electron incident from the left
  absorbs a phonon, which we will denote by the superscript
  $\mathrm{L,1}$;
\item Heating processes in which an electron incident from the left
  emits a phonon, denoted by the superscript $\mathrm{L,2}$;
\item Cooling processes in which an electron incident from the right
  absorbs a phonon, denoted by the superscipt $\mathrm{R,1}$; and
\item Heating processes in which an electron incident from the right
  emits a phonon, denoted by the superstcript $\mathrm{R,2}$.

\end{enumerate}
The power generated in a nanoscale junction is given by the sum of the
average power generated by those four processes, summed over all
possible vibrational modes:
\begin{equation}\label{totalpower}
W_\mathrm{tot}^\mathrm{avg.} = \sum_\mathrm{vib. modes}( 
\langle W^{\mathrm{R,2}} \rangle +
\langle W^{\mathrm{L,2}} \rangle -
\langle W^{\mathrm{R,1}} \rangle -
\langle W^{\mathrm{L,1}} \rangle)
\end{equation}

In addition to the above processes, cooling occurs due to dissipation
of energy in the bulk electrodes~\footnote{We assume here that at
  those biases when current-induced forces are large heating is small.
  This is generally true if dissipation into the bulk electrodes is
  efficient~\cite{yang:05}.}.  In order to evaluate the power
generated by each of those processes, we first need to consider the
full many-body Hamiltonian of the system:
\begin{equation}
H = H_\mathrm{el}+H_\mathrm{ion}+H_\mathrm{el-ion}
\end{equation}
Here, $H_\mathrm{el}$ is the electronic Hamiltonian (including
electron-electron effects), $H_\mathrm{ion}$ is the ionic Hamiltonian,
given by
\begin{equation}
H_{\mathrm{ion}} = \sum_{i=1}^N \frac{\vec{P}_i^2}{2M_i} +
\sum_{i,j}V_{\mathrm{ion}}(\vec{R}_i - \vec{R}_j ),
\end{equation}
and $H_{\mathrm{el-ion}}$ is the electron-ion interaction,
\begin{equation}
H_{\mathrm{el-ion}} = \sum_{i,j}V_{\mathrm{el-ion}}(\vec{r}_i-\vec{R}_j).
\end{equation}
$\vec{P}_i$, $M_i$ and $\vec{R}_i$ denote the momentum, mass
and position of the $i^{\mathrm{th}}$ ion (out of a total of $N$
ions), while $\vec{r}_i$ denotes the position of the
$i^{\mathrm{th}}$ electron.  

We assume that each ion executes a small vibration about its
equilibrium position $\vec{R}_j^0$ so that its diplacement is given by
$\vec{Q}_j = \vec{R}_j - \vec{R}_j^0$.  We can decouple these ionic
vibrations by introducing normal coordinates $\{q_{j\beta}\}$ so that
the $\alpha^{\mathrm{th}}$ component ($\alpha$ = $x$, $y$, $z$) of
$\vec{Q}_i$ is given by
\begin{equation}\label{decoupled}
(\vec{Q}_i)_\alpha = \sum_{j=1}^N \sum_{\beta=1}^3
A_{i\alpha,j\beta}q_{j\beta}.
\end{equation}
The coefficients $A_{i\alpha,j\beta}$ obey the orthonormality
relations $\sum_{i,\alpha}M_i A_{i\alpha,j\beta}$ $\times
A_{i\alpha,j'\beta'} = \delta_{j\beta,j'\beta'}$.  We now can solve
this problem in the usual way by introducing boson creation and
annihilation operators $b_{j\beta}^\dagger$ and $b_{j\beta}$ for the
$(j\beta)^\mathrm{th}$ mode; the creation and annihilation operators
obey the commutation relation $[b_{j\beta}, b_{j'\beta'}^\dagger] =
\delta_{j\beta,j'\beta'}$. With this transformation, the total ionic
Hamiltonian is just the sum of the Hamiltonians for each normal mode:
\begin{equation}
H_\mathrm{ion} = \sum_{j,\textrm{\hspace{1pt}}\beta}
(b_{j\beta}^\dagger b_{j\beta} +
\frac{1}{2} ) \omega_{j\beta}
\end{equation}

Next, much like in Section \ref{Sn}, we expand the field operator for
the electrons into a part that describes electrons incident from the
left, and a part that describes electrons incident from the right:
$\hat{\Psi} = \hat{\Psi}^\mathrm{L} + \hat{\Psi}^\mathrm{R}$.

We now express $H_{\mathrm{el-ion}}$ in terms of the fermionic and
bosonic creation and annihilation operators~\cite{chen:03:nl3}:
\begin{eqnarray}
H_\mathrm{el-ion} & = & \int \D \vec{r} \sum_i
V_\mathrm{el-ion}(\vec{r} - \vec{R}_i) \\
& = & \int \D \vec{r} \sum_i \vec{Q}_i \cdot \nabla_\mathbf{R}
V_\mathrm{el-ion}(\vec{r} - \vec{R}_i^0) \nonumber \\ 
& & \quad + O(Q^2) \\
& = & \sum_{\varphi,\chi}\sum_{E_1,E_2}\sum_{i\alpha,j\beta \in \mathrm{vib.}} 
\sqrt{ \frac{1}{2\omega_{j\beta}}} \nonumber \\
& & \label{Helion} \times A_{i\alpha,j\beta}
J_{E_1,E_2}^{i\alpha,\varphi\chi}a_{E_1}^{\varphi\dagger}a_{E_2}^{\chi}
(b_{j\beta} + b_{j\beta}^\dagger),
\end{eqnarray}
since we have $(\vec{Q}_i)_\alpha = \sum_{j\beta\in\mathrm{vib.}}
A_{i\alpha,j\beta}$ $ \sqrt{\frac{1}{2\omega_{j\beta}}}$ $(b_{j\beta}
+ b_{j\beta}^\dagger)$.  The electron-phonon coupling constant is
given by
\begin{eqnarray}
J_{E_1,E_2}^{i\alpha,\varphi\chi} & = & \int \D \vec{r} \int
\D \vec{K_\parallel}
\Psi^{\phi*}_{E_1}(\vec{r},\vec{K_\parallel}) \nonumber \\
& & \times \frac{\partial}{\partial R_\alpha}
V^{\mathrm{ps}}(\vec{r},\vec{R}_i^0)
\Psi^\chi_{E_2}(\vec{r},\vec{K_\parallel}),
\end{eqnarray}
where $V^{\mathrm{ps}}(\vec{r}, \vec{R}_i^0)$ denotes the
pseudopotential due to the $i^\mathrm{th}$ atomic core.  It is
interesting to note that, unlike the equilibrium case, in a
current-carrying wire the electron-phonon coupling constant depends on
two types of stationary states: left- and right-moving states.

\begin{figure}
  \centering
  \includegraphics[width=.4\textwidth]{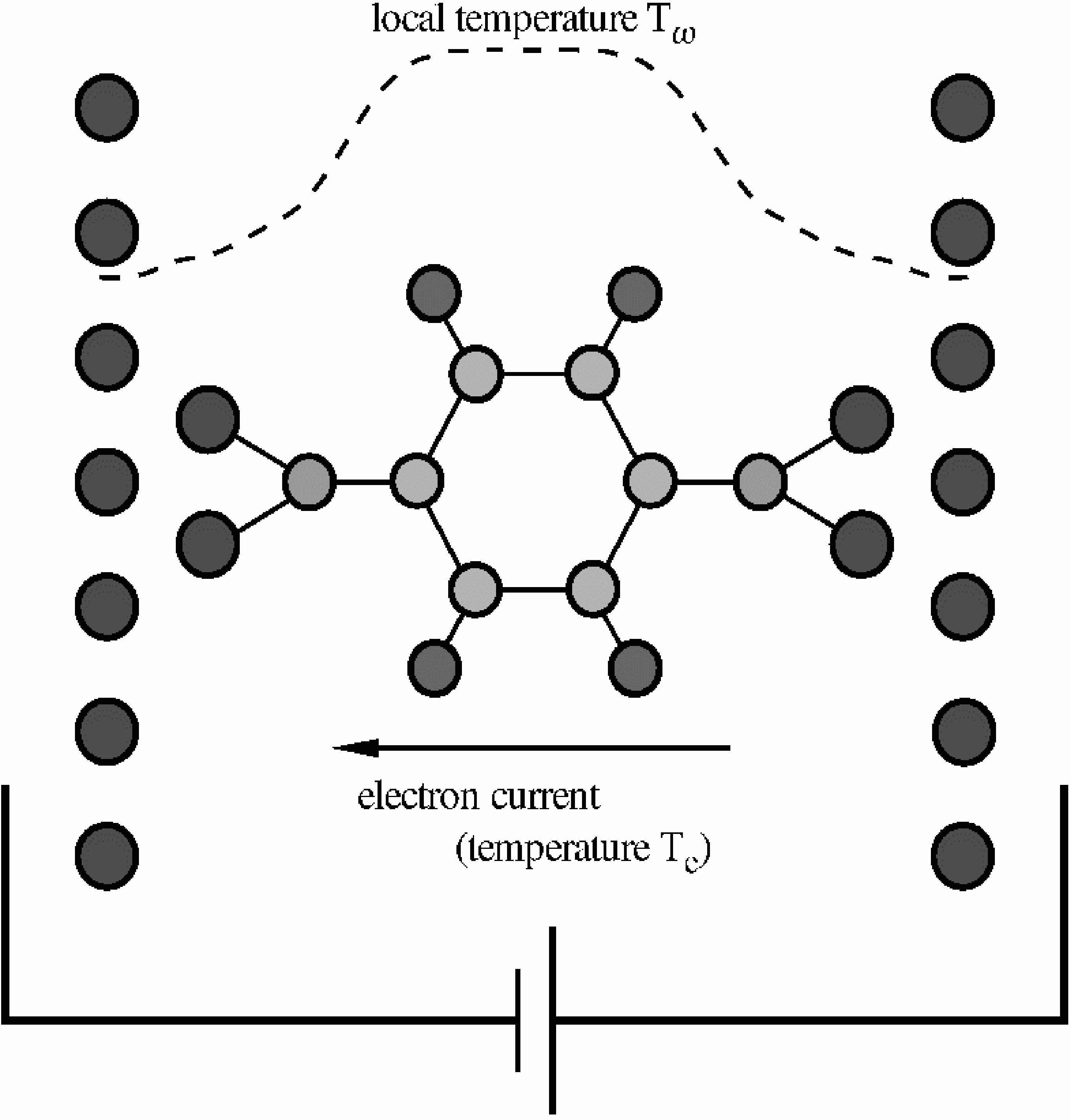}
  \caption{Molecular junction undergoing local heating as a result of
    transport.  The electron gas has a temperature $T_e$, and the
    phonons a temperature $T_\omega$.}
  \label{moljunct}
\end{figure}
We now assume that after continous exchange of energy, both the
electronic and phonon subsystems have reached a steady-state
temperature and therefore assign a temperature $T_e$ to the electron
gas, and a temperature $T_\omega$ to the phonons (see schematic in
figure~\ref{moljunct}), so that the statistics of the
$(j\beta)^\mathrm{th}$ phonon in the junction can be described by the
  Bose-Einstein distribution:
\begin{equation}
g_{j\beta} =
\frac{1}{ \E^{\omega_{j\beta}/k_\mathbf{B}T_\omega} - 1},
\end{equation}
where $T_\omega$ is the local temperature of the junction.

We can now use the Fermi's Golden Rule to infer rates of phonon
emission and absorption, and thus explicitly evaluate equation
(\ref{totalpower})~\cite{chen:03:nl3,montgomery:02,montgomery:03}.
Accordingly, the composite electron-phonon system's transition rate
from an initial state $|i\rangle$ to a final state $|f\rangle$ is
\begin{equation}\label{FGR}
R_{j\beta} = 2\pi | \langle i|
H_\mathrm{el-ion}^{j\beta} |f\rangle |^2
\delta(E_f - E_i - \omega_{j\beta} ).
\end{equation}
$E_i$ and $E_f$ denote the energy of the electron in the initial and
final state, respectively.  We are only considering the transition
rate for an electron interacting with the $(j\beta)^\mathrm{th}$ mode,
and so we have defined $H_\mathrm{el-ion}^{j\beta}$ to be the
$(j\beta)^\mathrm{th}$ term in the $H_\mathrm{el-ion}$ Hamiltonian
(see equation (\ref{Helion})).

Let us consider the case where an electron incident from the right
emits a phonon.  We can evaluate the statistical average of the
square of the matrix element of the Hamiltonian in equation
(\ref{FGR}) to obtain
\begin{eqnarray}\label{matrixelement}
& & \!\!\!\! \langle \Big| \langle i| H_\mathrm{el-ion}^{j\beta,\mathrm{R,2}}
|f\rangle
\Big|^2  \rangle = \\
& & \!\!\! \Bigg| \sum_{i\alpha}
\sqrt{ \frac{1}{2\omega_{j\beta}} } A_{i\alpha,j\beta}
J_{E-\omega_{j\beta},E}^{i\alpha,\mathrm{LR}}
\sqrt{ (1 + g_{j\beta} )
  f_E^\mathrm{R}(1-f_{E-\omega_{j\beta}}^\mathrm{L})
} \Bigg|^2. \nonumber
\end{eqnarray}
Note that the $1$ that is added to $g_{j\beta}$ corresponds to
spontaneous emission.

The power emitted by right-incident electrons is simply
{\setlength\arraycolsep{2pt}
\begin{eqnarray}
w_{j\beta}^\mathrm{R,2}(E_i,E_f) & = & (E_f - E_i) R_{j\beta} \\
& = & 2\pi \big|\langle i|
H_\mathrm{el-ion}^{j\beta} |f\rangle \big|^2 (E_f-E_i) \nonumber \\
& & \quad \times \delta(E_f - E_i - \omega_{j\beta}).
\end{eqnarray}}
$\!\!$The total power emitted by right-incident electrons to the
$(j\beta)^\mathrm{th}$ mode is the sum over all initial and final
states: {\setlength\arraycolsep{2pt}
\begin{eqnarray}
W_{j\beta}^\mathrm{R,2} & = & 2 \sum_{E_i} \sum_{E_f}
w_{j\beta}^\mathrm{R,2}(E_i,E_f) \\
& = & 2 \sum_{E_i} \sum_{E_f} 2\pi \big|\langle i|
H_\mathrm{el-ion}^{j\beta} |f\rangle \big|^2 (E_f-E_i) \nonumber \\
& & \quad \times \delta(E_f - E_i - \omega_{j\beta}) \\
& = & 2 \int \!\! \D E_i D_{E_i}^\mathrm{R}
\!\! \int \!\!\D E_f D_{E_f}^\mathrm{L}
2\pi \big|\langle i| H_\mathrm{el-ion} 
|f\rangle \big|^2 \nonumber\\
& & \quad \times(E_f-E_i) \delta(E_f - E_i - \omega_{j\beta}) \\
& = &  4\pi \!\! \int \!\! \D E_i
D_{E_i}^\mathrm{R} D_{E_i-\omega_{j\beta}}^\mathrm{L}
\big|\langle i| H_\mathrm{el-ion} |f\rangle \big|^2 
\omega_{j\beta},
\end{eqnarray}}
$\!\!$where we have multiplied by a factor of 2 due to spin
degeneracy. In taking the continuum limit, we have introduced
$D_E^\mathrm{R(L)}$, which is the partial density of states of
electrons moving to the right (left) with energy $E$.

Finally, the total power emitted by right-incident electrons to the
$(j\beta)^\mathrm{th}$ mode is therefore
{\setlength\arraycolsep{2pt}
\begin{eqnarray}
\langle W_{j\beta}^\mathrm{R,2} \rangle & = &
2\pi (1 +  g_{j\beta} )%
\int\!  \D E \bigg|\sum_{i\alpha} A_{i\alpha,j\beta}
J_{E-\omega_{j\beta}}^{i\alpha,\mathrm{LR}} \bigg|^2 \nonumber\\
& & \times f_E^\mathrm{R} (1-f_{E-\omega_{j\beta}})% 
D_E^\mathrm{R} D_{E-\omega_{j\beta}}^\mathrm{L}.
\end{eqnarray}}
$\!\!$Similar considerations apply to all other processes that
contribute to equation~(\ref{totalpower}).
\begin{figure}
  \centering
  \includegraphics[width=.4\textwidth,clip]{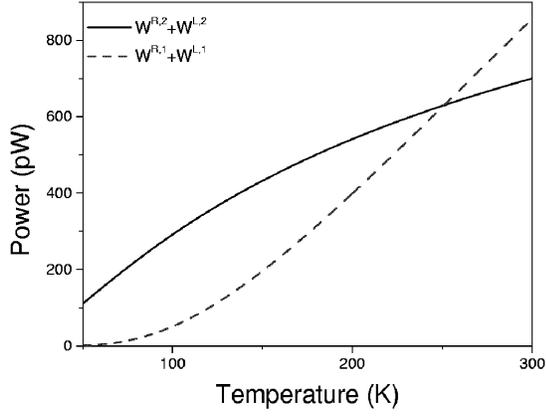}
  \caption{Absolute magnitude of the power due to electron-phonon
    interactions as a function of $T_\omega$.  The intersection of the
    two curves gives the steady-state temperature of the junction.}
  \label{heatintersect}
\end{figure}

The local temperature of the junction $T_\omega$ is evaluated (at a
fixed electronic temperature $T_e$) when the total
power~(\ref{totalpower}) is zero.  This is illustrated in Figure
\ref{heatintersect} where the magnitude of the sum of the heating
processes is plotted on the same graph as the magnitude of the sum of
the cooling processes as a function of $T_\omega$.  The point where
the two curves intersect is the steady-state temperature of the
nanojunction.

\begin{figure}
  \centering
  \includegraphics[width=.4\textwidth]{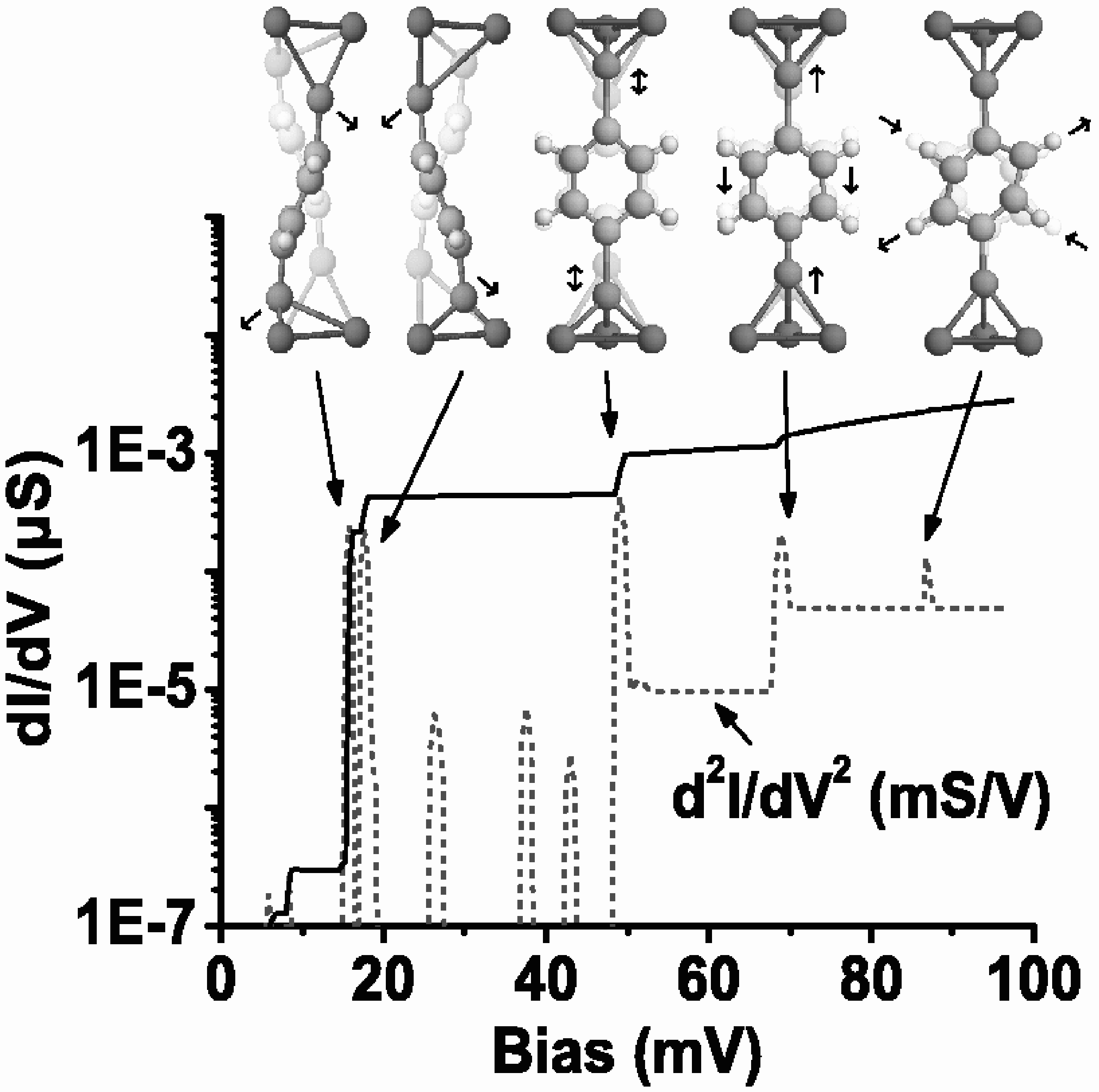}
  \caption{Steady-state temperature as a function of applied bias for
    two different contact geometries: a molecular junction, and a
    single gold atom contact.  A schematic of the contact geometry is
    shown as an inset for each case.  For both contact geometries, the
    nanojunction is not in thermal contact with the bulk electrodes.
    Reprinted with permission from \protect\cite{chen:03:nl3}.}
  \label{nocoupling}
\end{figure}
Let us now discuss two examples. We first neglect cooling due to
dissipation into the electrodes. Figure \ref{nocoupling} shows the
steady-state temperature as a function of applied bias for two
different nanojunctions, a gold point contact and a molecular
junction. We note that local heating occurs when a certain threshold
bias is reached. This is because a given vibrational mode cannot be
excited unless the incident electron is energetic enough to supply the
requisite amount of energy; that is, $V_\mathrm{Bias}^\mathrm{(crit.)}
= \min\{\omega_{i\alpha} \}$.  The onset bias for heating is in good
agreement with experimental observations in the gold point-contact
case~\cite{agrait:02}.  Once the threshold bias is reached, however,
the local temperature of the junction rises quickly, since cooling
processes cannot effectively compensate for heating processes. Note
that, in this case, very large temperatures can be reached at very
small biases.

\begin{figure}
  \centering
  \includegraphics[width=.4\textwidth]{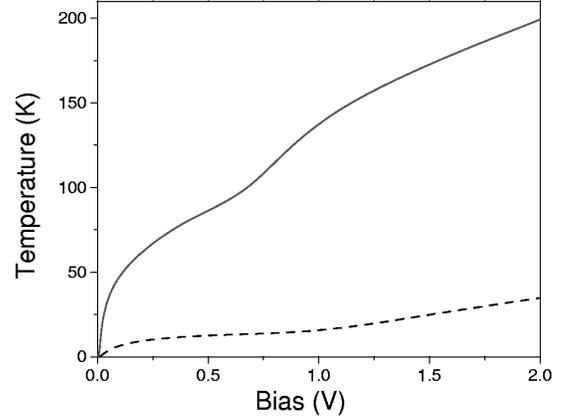}
  \caption{Steady-state temperature as a function of applied bias for
    the same geometries as in Figure \ref{nocoupling}, where the
    dashed line corresponds to the molecular junction, and the solid
    line corresponds to the gold point contact.  Dissipation into the
    bulk electrodes is taken into account.  Reprinted with permission
    from \protect\cite{chen:03:nl3}.}
  \label{coupling}
\end{figure}

This is in contrast to Figure \ref{coupling}, which depicts the same
process, but with an important difference: the nanojunction and
contacts are now assumed to be thermally coupled with each other, i.e.
energy can be dissipated into the bulk electrodes\index{cooling}. In
order to estimate this energy transfer we use the following expression
for elastic phonon scattering between a bulk material $\mathrm{A}$ in
contact with another bulk material $\mathrm{B}$ via a weak mechanical
link (the nanojunction)~\cite{patton:01}:
\begin{equation} \label{pattongeller}
I_\mathrm{th} = 4\pi K^2 \int\! \D \varepsilon \,
\varepsilon N_\mathrm{A}(\varepsilon) N_\mathrm{B}(\varepsilon)
[g(T_\mathrm{A},\varepsilon) - g(T_\mathrm{B}, \varepsilon)],
\end{equation}
where $g({T_\mathrm{A(B)},\varepsilon})$ is the Bose-Einstein
distribution at a temperature $T_\mathrm{A(B)}$ and energy
$\varepsilon$ and $N_\mathrm{A(B)}$ is the phonon spectral density of
states of surface $\mathrm{A(B)}$~\cite{patton:01}.  The weak
mechanical link is modeled via a harmonic oscillator with stiffness
$K$.

Applying equation (\ref{pattongeller}) to our case we obtain the
results of Figure \ref{coupling}~\cite{chen:03:nl3}.  Notice that the
bias scale in Figure \ref{coupling} is much different than the one in
Figure \ref{nocoupling}, i.e. the majority of heat generated in the
junction is dissipated into the bulk electrodes. However, in real
devices poor thermal contact between the junction and the electrodes
can actually occur, as can localized phonon modes within the junction
which have low coupling with the continuum of modes of the bulk
electrodes~\cite{montgomery:03}.  Such instances can lead to very
large local temperatures in the junction with consequent structural
instabilities~\cite{chen:03:nl3,montgomery:03,yang:05}.  Such instabilities
have been actually observed in metallic point contacts~\cite{smit:04} and
may be responsible for the low yield in fabricating molecular
junctions~\cite{zhitenev:04}.

\section{Inelastic Conductance}
\index{inelastic conductance}
\index{conductance!inelastic}

\begin{figure}
  \centering \includegraphics[width=.4\textwidth]{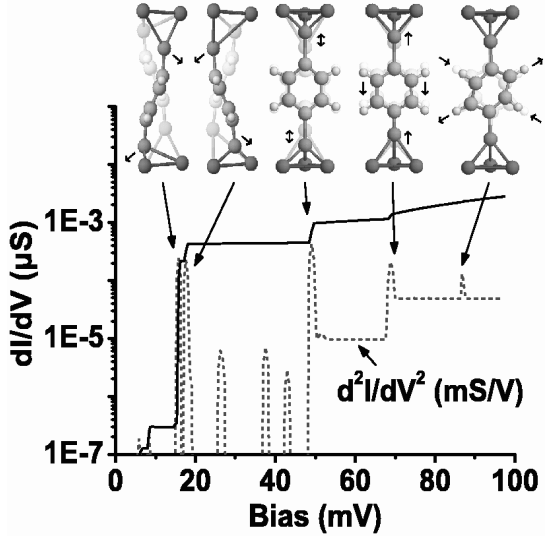}
  \caption{Magnitude of the differential conductance through a
    molecular junction made of a benzene-dithiolate molecule.  The
    dashed line gives the derivative of the differential conductance,
    while the insets illustrate the major longitudinal phonon modes
    Reprinted with permission from \cite{chen:04}.  Copyright 2004
    American Chemical Society.}
  \label{fig:InCond}
\end{figure}

In addition to local heating, electron-phonon coupling can yield a lot
of information on the internal structure of nanoscale junctions:
discontinuities in the conductance can occur when the energy of
incident electrons becomes large enough to excite different
vibrational modes of the junction.

In order to calculate the inelastic current, let us consider the case
of an electron incident from the right. By treating the
electron-phonon interaction as a perturbation, the total
wavefunctions of the (electron plus phonon) system can be written in
terms of the states $|\Psi_E^\mathrm{R};n_{i\alpha}\rangle =
|\Psi_E^\mathrm{R}\rangle \otimes |n_{i\alpha}\rangle$.  The
wavefunction of the system, including electron-phonon interactions is
therefore (to first order): $|\Phi_E^\mathrm{R}; n_{i\alpha}\rangle =
|\Psi_E^\mathrm{R}; n_{i\alpha}\rangle + |\delta
\Psi_E^\mathrm{R};n_{i\alpha}\rangle$, where $|\delta
\Psi_E^\mathrm{R};n_{i\alpha}\rangle$ is the leading-order change in
the wavefunction due to electron-phonon interaction.  We can then
calculate the leading correction to the total wavefunction:
\begin{eqnarray}
\lefteqn{ |\delta\Psi_E^\mathrm{R};n_{j\beta}\rangle
= \lim_{\varepsilon \to 0^+} \sum_{\phi'=\mathrm{L,R}} \sum_{j',\beta'} \int\!
\D E' D_{E'}^{\phi'} } \nonumber \\
& & \times \frac{ \langle \Psi_{E'}^{\phi'};n_{j'\beta'} | H_\mathrm{el-ion} | 
\Psi_E^\mathrm{R};n_{j\beta} \rangle
|\Psi_{E'}^{\phi'};n_{j'\beta'} \rangle }{%
\varepsilon(E,n_{j\beta}) - \varepsilon(E',n_{j'\beta'}) - \I \varepsilon }.
\end{eqnarray}
Once again, $D_{E}^\mathrm{L(R)}$ is the partial density of states for
electrons incident from the left (right).  $\varepsilon(E,n_{j\beta})
= E + (n_{j\beta} + \frac{1}{2} )\omega_{j\beta}$ is the energy of
state $|\Psi_E^{\phi};n_{j\beta}\rangle$.\footnote{Since the
  electron-phonon coupling is small for the systems considered here,
  we assume the energy $E$ is the unperturbed electronic energy, i.e.
  the elastic correction to the electronic energy is negligible.}

Using the identity $\lim_{\varepsilon \to 0} \frac{1}{z-i\varepsilon}
= P(\frac{1}{z}) + i\pi\delta(z)$, where $P(x)$ denotes the principle
value of $x$, together with the matrix element that leads to equation
(\ref{matrixelement}), we obtain
\begin{widetext}
\begin{eqnarray}
|\delta\Psi_E^\mathrm{R};n_{j\beta}\rangle  & = & 
\I \pi \sum_{i\alpha} \sqrt{\frac{1}{2\omega_{j\beta}} }%
A_{i\alpha, j\beta} \nonumber \bigg[D_{E+\omega_{j\beta}}^\mathrm{L}%
\sqrt{ g_{j\beta} %
  f_E^\mathrm{R}( 1-f_{E+\omega_{j\beta}}^\mathrm{L} ) }
J_{E+\omega_{j\beta}}^{i\alpha,\mathrm{LR}}%
|\Psi_{E+\omega_{j\beta},E}^\mathrm{L};n_{j\beta}-1 \rangle \nonumber \\
& & + D_{E-\omega_{j\beta}}^\mathrm{L}%
\sqrt{( 1+ g_{j\beta} )%
  f_E^\mathrm{R}( 1-f_{E-\omega_{j\beta}}^\mathrm{L} ) }
J_{E-\omega_{j\beta}}^{i\alpha,\mathrm{LR}}%
|\Psi_{E-\omega_{j\beta},E}^\mathrm{L};n_{j\beta}+1 \rangle
\bigg]
\end{eqnarray}
\end{widetext}
The analagous expression for
$|\delta\Psi_E^\mathrm{L};n_{j\beta}\rangle$ is obtained by simply
interchanging all $\mathrm{R}$'s with $\mathrm{L}$'s.

As an axample let us assume that the electronic temperature $T_e$ is
equal to zero.  In that case, for an external bias $V = E_\mathrm{FL}
- E_\mathrm{FR}$, only normal modes with energies $\omega_{j\beta} <
V$ can be excited.  (Again, we are assuming the left electrode is
positively biased.)  Furthermore, if we assume negligible local
heating, and that the few excited phonons decay on a short time scale,
then we have $g_{j\beta} = 0$.  In this case, the inelastic
contribution to the current is:
\begin{eqnarray}
\delta I & = & -\I \int_{E_\mathrm{FL}}^{E_\mathrm{FR}} \!
\D E \int\! \D \vec{R} \int\! \D \vec{K_\parallel} \nonumber \\
& & \big[ (\delta\Psi_E^\mathrm{R})^*\partial_z\delta\Psi_E^\mathrm{R} -
\partial_z(\delta\Psi_E^\mathrm{R})^*\delta\Psi_E^\mathrm{R} \big].
\end{eqnarray}

In Figure \ref{fig:InCond} we plot the inelastic conductance of a
molecular junction. It is evident that only specific modes with large
longitudinal component (with respect to the direction of current flow)
dominate the inelastic conductance, while modes with large transverse
component contribute negligibly. In addition, it is shown in Ref.
\cite{chen:04} that inelastic current-voltage characteristics are
quite sensitive to the structure of the contact between the molecule
and the electrodes thus providing a powerful tool to extract the
bonding geometry in molecular wires.

\section{Conclusions}

We have presented a review of several current-induced effects in
nanoscale conductors and their description at the atomic level. These
effects provide a wealth of information on the transport properties of
atomic and molecular junctions beyond the value of the average
current. In addition, their understanding is paramount to the possible
application of nanoscale systems in electronics.
\\
\\
\begin{acknowledgments}
\noindent One of us (MD) is indebted to the students and research
associates who have worked with him over the past years on the issues
described in this review: M. Zwolak, J. Lagerqvist, Y.-C. Chen, M.
Chshiev and Z.  Yang.  We are also thankful for illuminating
discussions with N.D.  Lang and T.N. Todorov.  We acknowledge support
from the NSF Grant No.  DMR-01-33075.
\end{acknowledgments}

% BibTeX users please use
%\bibliographystyle{LNiP}
%\bibliography{bushong}
%
% Non-BibTeX users please follow the syntax
% the syntax of "referenc.tex" for your own citations
% \input{referenc}

%%%%%%%%%%%%%%%%%%%%%%%%%%%%%%%%%%%%%%%%%%%%%%%%%%%%%%%%%%%%%%%%%%%%%%  }

%%%%%%%%%%%%%%%%%%%%%%%%%%%%%%%%%%%%%%%%%%%%%%%%%%%%%%%%%%%%%%%%%%%%%%

\printindex
\end{document}